\documentclass[onecolumn,prl, preprint, superscriptaddress]{revtex4-2}
\usepackage{bm}
\usepackage[colorlinks=true,linkcolor=blue,citecolor=blue]{hyperref}
\usepackage{times}
\usepackage{amsmath}
\usepackage{amssymb}
\usepackage{amsthm}
\usepackage{amsfonts}
\usepackage{enumerate}
\usepackage{latexsym}
\usepackage{ifpdf}
\newcommand{\beq}{\begin{equation}}
\newcommand{\eeq}{\end{equation}}
\usepackage{graphicx}
\usepackage{makeidx}
\usepackage{color}

\usepackage[version=4]{mhchem}
\usepackage{textcomp,mathcomp}

\begin{document}

\title{Interfacial reconstruction effects in insulating double perovskite Nd$_2$NiMnO$_6$/SrTiO$_3$ and Nd$_2$NiMnO$_6$/NdGaO$_3$ thin films}

\author {Nandana Bhattacharya}
\email{nandanab@iisc.ac.in}
\affiliation  {Department of Physics, Indian Institute of Science, Bengaluru  560012, India}

\author {Ranjan Kumar Patel}
\altaffiliation{Currently at Department of Electrical and Computer Engineering,
Rutgers University, Piscataway, NJ 08854, USA}
\affiliation  {Department of Physics, Indian Institute of Science, Bengaluru 560012, India}
\author {Siddharth Kumar}
\altaffiliation{Currently at Department of Electrical and Computer Engineering,
Rutgers University, Piscataway, NJ 08854, USA}
\affiliation{Department of Physics, Indian Institute of Science, Bengaluru 560012, India}
\author {Prithwijit Mandal}
\affiliation  {Department of Physics, Indian Institute of Science, Bengaluru 560012, India}
\author {Jyotirmay Maity}
\affiliation{Department of Physics, Indian Institute of Science, Bengaluru 560012, India}
\author{Christoph Klewe}
\affiliation  {Advanced Light Source, Lawrence Berkeley National Laboratory, Berkeley, California 94720, USA}
\author{Zhan Zhang}
\affiliation {Advanced Photon Source, Argonne National Laboratory, Lemont, Illinois 60439, USA}
\author{Hua Zhou}
\email{hzhou@anl.gov}
\affiliation {Advanced Photon Source, Argonne National Laboratory, Lemont, Illinois 60439, USA}

\author {Srimanta Middey}
\email{smiddey@iisc.ac.in}
\affiliation  {Department of Physics, Indian Institute of Science, Bengaluru 560012, India}

\clearpage

\begin{abstract}

Ferromagnetic insulating (FMI) double perovskite oxides (DPOs) $A_2BB'$O$_6$ with near-room-temperature Curie temperatures are promising candidates for ambient-temperature spintronics applications. 
To realize their potential, epitaxial stabilization of DPO films and understanding the effect of multiple broken symmetries across the film/substrate interface are crucial.
This study investigates ultrathin films of the FMI Nd$_2$NiMnO$_6$ (NNMO) grown on SrTiO$_3$ (STO) and NdGaO$_3$ (NGO) substrates. By comparing growth on these substrates, we examine the influence of polarity and structural symmetry mismatches, which are absent in the NGO system. The interface exhibits immeasurable resistance in both cases.
Using synchrotron X-ray diffraction, we show that films have three octahedral rotational domains because of the structural symmetry mismatch with the STO substrate. Furthermore, our coherent Bragg rod analysis of specular X-ray diffraction reveals a significant modification of the out-of-plane lattice parameter within a few unit cells at the film/substrate interface and the surface. This arises from polarity compensation and surface symmetry breaking, respectively. These structural alterations influence the Mn orbital symmetry, a dependence that we further confirm through X-ray linear dichroism measurements. 
Since the ferromagnetism in insulating DPOs is mediated by orbital-dependent superexchange interactions [Phys. Rev. Lett. {\bf 100}, 186402 (2008)], our study provides a framework for understanding the evolution of magnetism in ultrathin geometry.

\end{abstract}

\maketitle

\section{Introduction}

Ferromagnetic insulators (FMIs) are rapidly emerging as a promising class of materials for modern technology, offering minimal power loss and enabling the development of advanced, energy-efficient spintronic devices~\cite{Wolf:2001p1488,Brataas:2020p885,Emori:2021p020901}. To realize the full potential of FMI-based technology, two key requirements must be met: a ferromagnetic transition temperature ($T_\mathrm{c}$) exceeding room temperature in thin film geometry and a close lattice match between the film and widely available substrates. 
Due to advancements in oxide thin-film growth techniques, significant efforts are underway to meet these requirements using transition metal perovskite oxides ($AB$O$_3$)~\cite{Lee:2010p954, Lu:2022p7580, Li:2023p3638, Meng:2018p2873, Choi:2014p14836, Liu:2023v4, Li:2020p1901606}.  
The ferromagnetic double perovskite oxides (DPOs)  family with the general formula A$_2$\emph{BB}'O$_6$ (\emph{B}, \emph{B}' are the transition metals) is another promising platform in meeting these criteria with additional benefits ~\cite{Vasala:2015p1, SahaDasgupta:2020p014003}. 
About a thousand DPOs have already been synthesized in bulk form, and many show magnetic transition temperatures higher than 300 K ~\cite{Ray:2011p47007,Deng:2024p20239, Kang:2023p18474, Borges:1999p445}. Furthermore, machine learning-based methods are being used to design and synthesize new DPOs with the desired functionality from a much broader palette of elements from the periodic table~\cite{Saha:2024p023001, Halder:2019p084418,Tang:2022p15301, Guo:2024p6103}. Realizing DPO-based spintronic devices hinges on understanding the interface between FMI DPOs and insulating substrates.  Broken symmetry at this interface can drive diverse structural, electronic, orbital, and magnetic reconstructions~\cite{Zubko:2011p141, Ramesh:2019p257, Hwang:2012p103, Huang:2018p1802439, Schlom:2007p589, Schlom:2008p2429, Chakhalian:2014p1189}, potentially affecting the ferromagnetic transition temperature ($T_\mathrm{c}$) significantly~\cite{Das:2008p186402}.

Akin to silicon's dominance in semiconductor technology, SrTiO$_3$ (STO) is a popular choice as a substrate in oxide electronics due to its high dielectric constant, excellent lattice compatibility with numerous compounds with perovskite structure, and robust thermal and chemical stability ~\cite{Muller:1979p3593}. In this work, we investigate the interfacial effects between an FMI DPO Nd$_2$NiMnO$_6$ (NNMO) film and STO (001) substrates at room temperature. Bulk NNMO has a monoclinic structure and undergoes a paramagnetic to ferromagnetic transition at $T_\mathrm{c}$ of 200 K~\cite{Pal:2019p045122}.
Along the pseudocubic [001] direction, NNMO has alternating [Ni$_{0.5}$Mn$_{0.5}$O$_2$]$^{-1}$  and [NdO]$^{+1}$ layers, resulting in a polar mismatch with the nonpolar STO, which consists of alternating [SrO]$^0$ and [TiO$_2$]$^0$ layers [Fig.~\ref{Fig1}(a)]. Furthermore, the octahedral rotation patterns of the two materials differ: $a^-b^+c^-$ (in Glazor notation)  for NNMO and $a^0a^0a^0$ (no octahedral rotation) for STO [Fig.~\ref{Fig1}(a)]. While recent studies have examined the influence of polar catastrophe on electronic properties~\cite{Spurgeon:2018p134110, DeLuca:2022p2203071, Bhattacharya:2025p176201}, the interplay between polar and structural symmetry mismatches and their consequent impact remains unexplored. This work aims to unravel this intricate interplay by employing synchrotron X-ray diffraction~\cite{May:2010p014110, Disa:2020p1901772} and X-ray linear dichroism (XLD)~\cite{Chakhalian:2007p318} techniques.

\par In this article, we investigate NNMO films of thicknesses 5, 10, and 20 uc [uc= unit cell in pseudocubic ($\mathrm{pc}$) notation] grown on STO (001) and NdGaO$_3$ (NGO) (110)$_\mathrm{or}$ [(001)$_\mathrm{pc}$] [or= orthorhombic settings] substrates. By comparing the behavior of NNMO films on these two substrates, we can isolate the effects of structural and polar mismatches, as the NNMO/NGO interface is devoid of both [Fig.~\ref{Fig1}(a)]. Considering the lattice constant of bulk NNMO [$a$= 5.414 {\AA}, $b$ = 5.463 {\AA}, $c$ = 7.669 {\AA}]~\cite{Pal:2019p045122}, the epitaxial strain is calculated to be +0.4\%, and +1.5\% for NNMO film on NGO, and STO, respectively.
Synchrotron X-ray diffraction reveals a complex domain structure in NNMO/STO films, with three distinct octahedral rotational domains, in contrast to the single-domain behavior of NNMO/NGO films. Our 1D coherent Bragg rod analysis (COBRA) of the specular X-ray crystal truncation rod demonstrates an enhancement of the out-of-plane lattice parameter $c_\mathrm{pc}$ at few interfacial layers as a result of emerging Mn$^{3+}$ states to compensate polarity for the film on STO. Interestingly, we found an enhanced $c_\mathrm{pc}$ near the surface layers of the film on both STO and NGO systems, which arises due to the surface effect. Furthermore, X-ray linear dichroism (XLD) experiments unveil a thickness-dependent competition between tensile strain and surface symmetry breaking, which significantly influences the orbital symmetry of Mn cations in NNMO/STO films.

\section{Methods}

NNMO films were grown in layer-by-layer fashion on TiO$_{2}$ terminated cubic (\emph {Pm-3m}) STO (001) substrates and GaO$_{2}$ terminated orthorhombic (\emph{Pbnm}) NGO (110)$_\mathrm{or}$ [(001)$_\mathrm{pc}$] substrates using a Neocera pulsed laser deposition (PLD) system. During deposition, a KrF excimer laser operating at $\lambda$ = 248 nm with fluence of 2 J/cm$^2$ and a repetition rate 2 Hz was used. For films on both NGO and STO substrates, the growth conditions were maintained similar, with a deposition temperature of 750 $^\circ$C and a dynamic oxygen pressure of 150 mTorr maintained during deposition. The films were then annealed post-growth at a 500 Torr oxygen pressure at the deposition temperature. The growth was monitored with in-situ reflection high energy electron diffraction (RHEED) imaging. Fig.~\ref{Fig1}(b) shows the variation of intensity oscillation of the specular spot (00) during the growth of an NNMO film on STO substrate, which testifies a layer-by-layer growth with unit cell precision. Observation of streaky pattern in the RHEED images of the films obtained post-cooling to room temperature confirms smooth surface morphology [inset Fig.~\ref{Fig1}(b)]. Additional half-order diffraction streaks, observed in the film RHEED (denoted by arrows in inset Fig.~\ref{Fig1}(b)), arise due to in-plane doubling of the unit cell and are indicative of the orthorhombic/monoclinic symmetry. The NNMO films were characterized by x-ray reflectivity (XRR) and x-ray diffraction (XRD) measurements using a Rigaku Smartlab X-ray diffractometer. The fitting of XRR data using GenX program~\cite{Bjorck:2007p1174} [Fig.~\ref{Fig1}(c)] confirms a sharp film/substrate interface with cationic intermixing limited within one uc.  We further employed wire-bonded contacts to examine the electrical transport properties of the interface and observed no measurable conductivity.

Synchrotron X-ray diffraction measurements were performed at room temperature using an incident photon energy of 15.5 keV at the 33-ID-D beamline of the Advanced Photon Source (APS), USA. A pixel array area detector, Dectris PILATUS 100K, was used to detect and map the two-dimensional image of the diffraction spot for individual $L$-scan step. The background correction from fluorescence and diffuse scatterings was subtracted, and geometric factor corrections were performed to obtain the absolute intensity from the crystal truncation rod (CTR) diffraction alone ~\cite{Disa:2020p1901772, Schleputz:2011p73}.

X-ray absorption spectroscopy (XAS) with linearly polarized light (in-plane and out-of-plane) (angle of incidence $\sim$ 15$^{\circ}$), performed at beamline 4.0.2 of the Advanced Light Source (ALS), Lawrence Berkeley National Laboratory, USA 
was employed to probe NNMO films of various thicknesses on NGO and STO substrates. The difference between the in-plane and out-of-plane XAS spectra, known as X-ray linear dichroism (XLD), probes the orbital character of transition metal ions ~\cite{Chakhalian:2007p318}. XAS spectra were recorded in total electron yield (TEY) mode at the Mn and Ni $L_{3,2}$ edges at 300 K. The effective probing depth for the TEY mode is $\sim$ 9-10 nm~\cite{Stohr:2006book}, which is sufficient to probe the electronic structure of the film-substrate interface for all our film thicknesses ($\sim$ 2-8 nm).

\section{Results and discussions}

\subsection{Structural characterization of NNMO films}

\begin{figure*} 
	
    \includegraphics[width=0.83\textwidth] {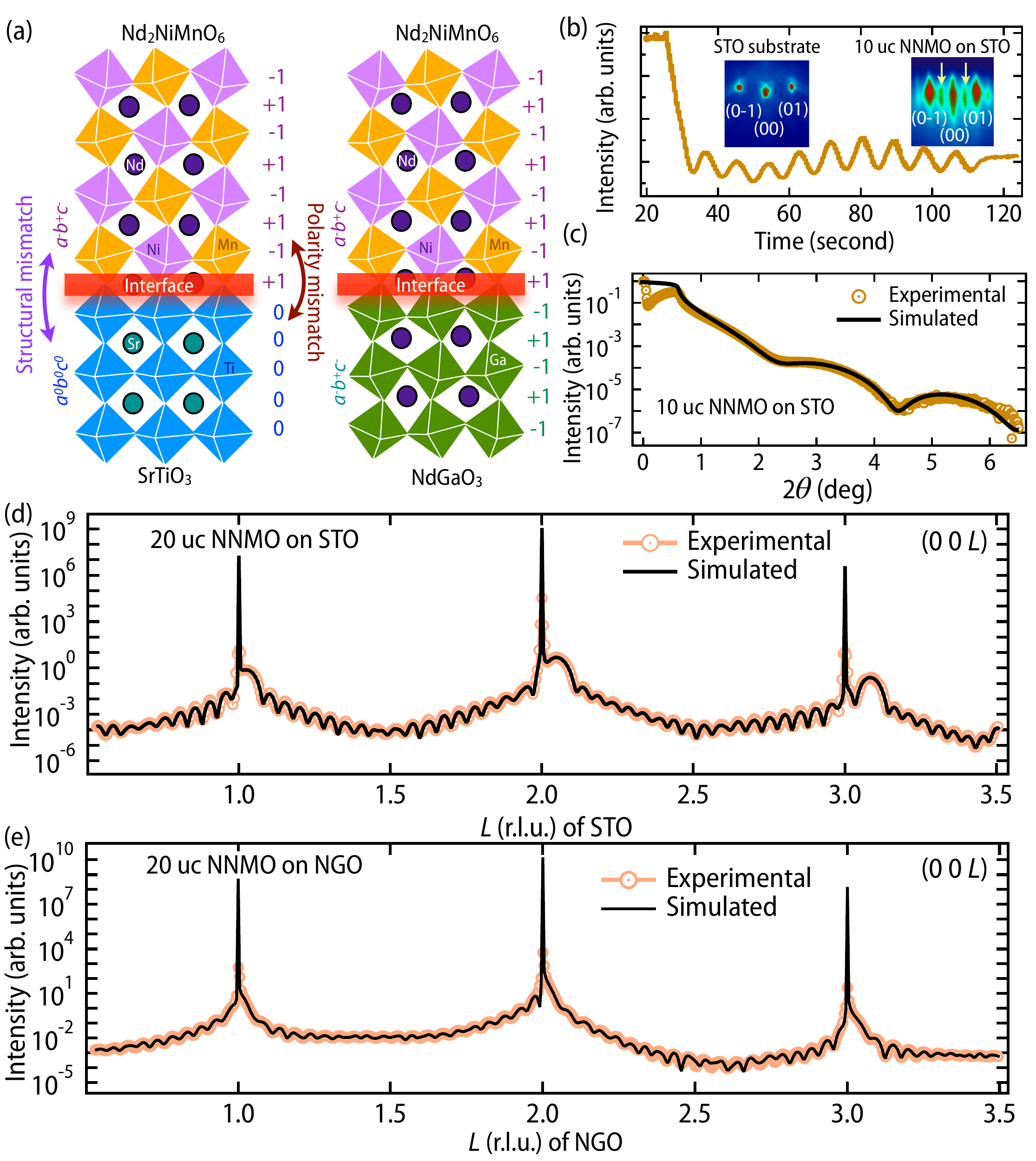}
	\caption{\label{Fig1}{(a) Schematic depicting the structural and polarity mismatch between NNMO and STO (left) and, NNMO and NGO with none of these mismatches (right) (b) RHEED oscillations for 10 uc NNMO on STO; Inset : RHEED image of pristine STO substrate (left), RHEED image for 10 uc film on STO after post annealing and cooling to room temperature (right). Yellow arrows indicate half-order diffraction.
    (c) Experimental XRR data for 10 uc NNMO on STO along with fitted curve using GenX. Diffraction pattern along (00$L$)$_\mathrm{pc}$ CTR and corresponding simulated curve for (d) 20 uc NNMO on STO (e) 20 uc NNMO on NGO.}}
\end{figure*}

The reciprocal space of an epitaxial system comprises crystal truncation rods (CTRs) characterized by fixed \emph{H} and \emph{K} indices and a continuous range of \emph{L} values. These CTRs intersect the Ewald sphere, satisfying Bragg's diffraction condition, and collectively construct the diffraction pattern~\cite{Kaganer:2007p245425, Yuan:2018p5220,Disa:2020p1901772}. CTR and half-order diffraction measurements [$L$ scan for several $(H K L)$ reflections] were performed to investigate the global crystallographic periodicity and octahedral rotation patterns. The underlying NGO and STO substrates have defined all crystallographic directions in this article.  We begin our discussion with the results of the specular (00$L$) CTR of the 20 uc NNMO films grown on NGO and STO substrates, as depicted in Fig.~\ref{Fig1}(d) and (e). The broad film peaks observed near the sharp substrate peaks at integral $L$ values confirm the single-crystalline nature of the films. While the NNMO film on NGO, due to a small tensile strain, exhibits overlapping peaks with the substrate, the presence of thickness fringes in both cases testifies a flat film/substrate interface. We also simulated the (00$L$) diffraction patterns using a 1D COBRA algorithm~\cite{Zhou:2012p195302,Disa:2020p1901772}. The simulated patterns, also shown in Fig.~\ref{Fig1}(d) and (e), closely match the experimental data. More details about the simulation and the outcome of the fitting are presented later in this article.

\begin{figure*}[h!]
	\vspace{-1pt}
	\hspace{0pt}
	\includegraphics[width=0.85\textwidth] {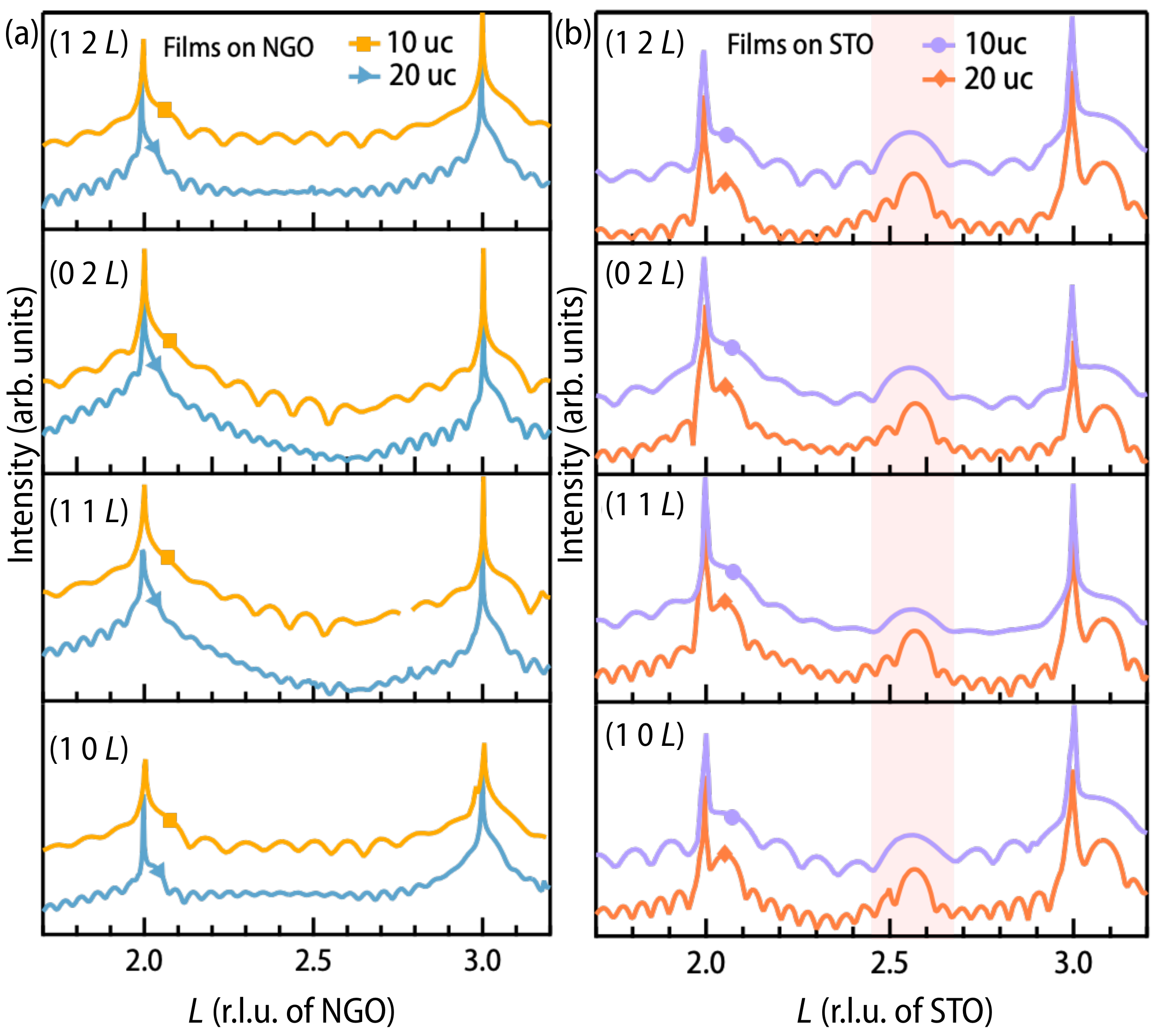}
	\caption{\label{Fig2}{XRD about non-specular integral CTRs (1 2 $L$)$_\mathrm{pc}$, (0 2 $L$)$_\mathrm{pc}$, (1 1 $L$)$_\mathrm{pc}$ and (1 0 $L$)$_\mathrm{pc}$ for 10 and 20 uc films on (a) NGO and (b) STO. Prominent additional half order peaks are highlighted with pink shade in (b) and are present exclusively for films on STO. The 10 uc spectra in each panel have been vertically shifted for visual clarity.}}
\end{figure*}

Interfacial polarity and symmetry mismatches can lead to the formation of non-perovskite phases characterized by various types of oxygen vacancy ordering, such as brownmillerite, La$_{2}$Ni$_{2}$O$_{5}$-like, or Ca$_{2}$Mn$_2$O$_5$-like phases, etc~\cite{Stolen:2006p429,Jeen:2013p1057,Middey:2014p6819,Andersen:2018p5949,Meyer:2016p1500201,Tsuji:2017p1864,Kim:2014p14646}.
The absence of half-order peaks in the (00\emph{L}) diffraction [Fig.~\ref{Fig1}(d), (e)] excludes the formation of a brownmillerite phase for NNMO films on STO~\cite{Jeen:2013p1057}. To investigate the possibilities of other  oxygen vacancy-ordered scenarios, we examined off-specular integral rods ($H$ $K$ $L$) where at least one of the $H$ and $K$ is a non-zero integer. 
While the films on NGO exhibited diffraction peaks only at integral $L$ values in these off-specular scans [Fig.~\ref{Fig2}(a)], those on STO showed additional peaks at half-integer $L$ values [Fig.~\ref{Fig2}(b)]. 
Interestingly, the intensity of these half-order peaks are of comparable order of magnitude to the film intensity. If this arises due to oxygen vacancy ordered phases, a significant alteration in oxygen stoichiometry would be necessary throughout the bulk volume of the film. This would lead to a considerable change in cationic valence to satisfy charge conservation. However, our XAS and electron energy loss spectroscopy (EELS) measurements rule out significant oxygen vacancy concentrations in the films but rather show the variation of Mn oxidation states near the interface only~\cite{Bhattacharya:2025p176201}.  Thus, a more plausible scenario explaining these peaks may emanate from the global octahedral symmetry of the films. To further explore that, we have investigated the octahedral rotational patterns of the films using half-order Bragg rods. 

\subsection{Determination of global octahedral rotational pattern (ORP) using half-order Bragg rods} 

 \begin{figure*}[ht!]
	\centering
	{~} \vspace{-2pt}
	{~} \hspace{0pt}
	\includegraphics[width=1.0 \textwidth] {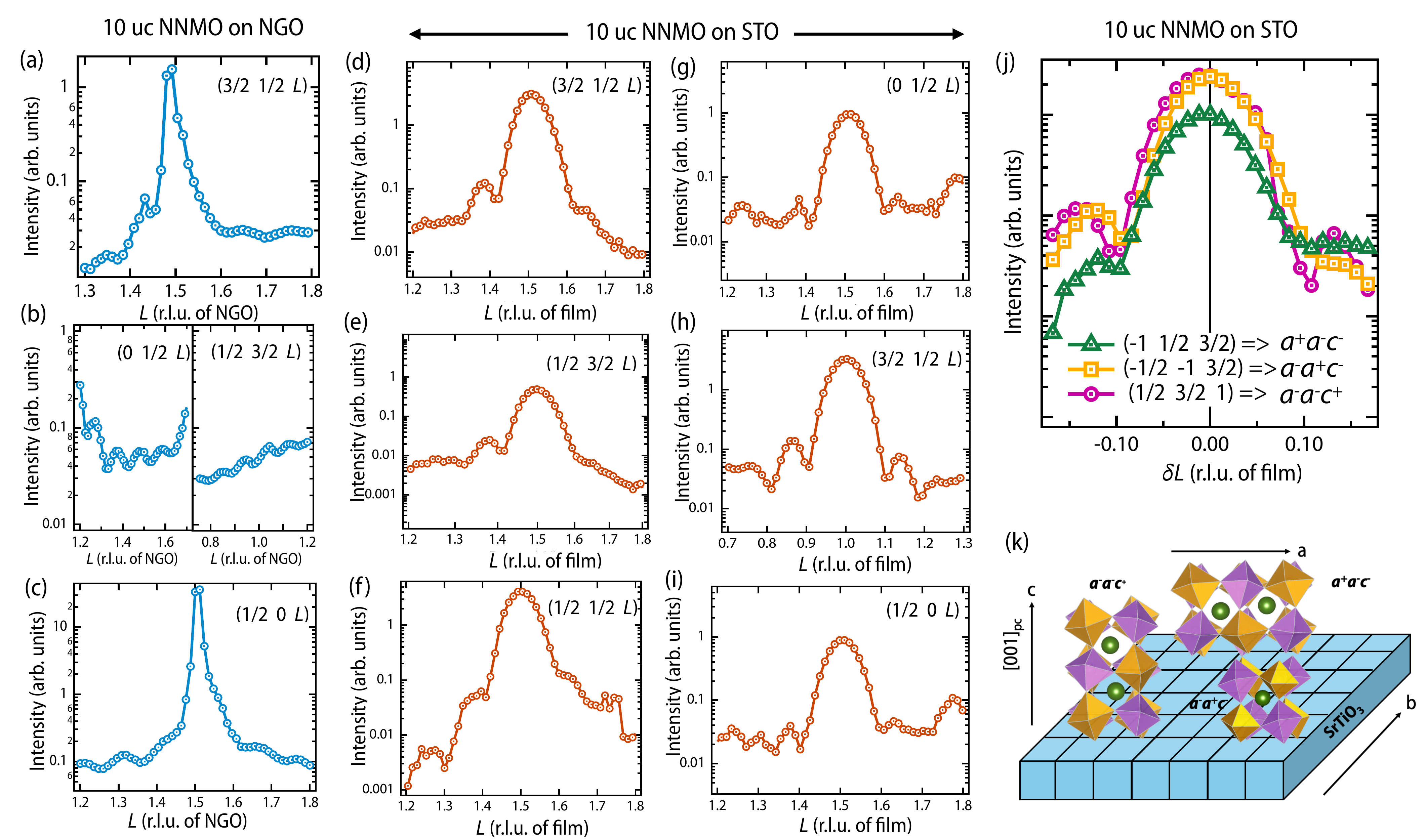}
	\caption{\label{Fig3}{Half-order diffraction highlighting the peaks (a) ($\frac{3}{2}$  $\frac{1}{2}$  $\frac{3}{2}$)$_\mathrm{pc}$ (b) (0  $\frac{1}{2}$  $\frac{3}{2}$)$_\mathrm{pc}$ (left) ($\frac{1}{2}$ $\frac{3}{2}$ 1)$_\mathrm{pc}$ (right) (c) ($\frac{1}{2}$  0  $\frac{3}{2}$)$_\mathrm{pc}$ for a 10 uc NNMO film on NGO. Half order diffraction highlighting the peaks (d) ($\frac{3}{2}$  $\frac{1}{2}$  $\frac{3}{2}$)$_\mathrm{pc}$ (e) ($\frac{1}{2}$  $\frac{3}{2}$  $\frac{3}{2}$)$_\mathrm{pc}$ (f) ($\frac{1}{2}$  $\frac{1}{2}$  $\frac{3}{2}$)$_\mathrm{pc}$ (g) (0  $\frac{1}{2}$  $\frac{3}{2}$)$_\mathrm{pc}$ (h) ($\frac{3}{2}$  $\frac{1}{2}$  1)$_\mathrm{pc}$ and (i) ($\frac{1}{2}$  0  $\frac{3}{2}$)$_\mathrm{pc}$ for a 10 uc NNMO film on STO. Since half order peaks for cubic STO are absent, the x-axis in these plots for film on STO correspond to the $L$ of the film. (j) Half order peaks about ($H$', $K$', $L$'), $\delta$\emph{L} = $L$ - $L$' corresponding to \emph{a$^-$a$^-$c$^+$} [($\frac{1}{2}$  $\frac{3}{2}$ 1)$_\mathrm{pc}$], \emph{a$^-$a$^+$c$^-$} [(-$\frac{1}{2}$  -1  $\frac{3}{2}$)$_\mathrm{pc}$] and \emph{a$^+$a$^-$c$^-$} [(-1  $\frac{1}{2}$  $\frac{3}{2}$)$_\mathrm{pc}$]. (k) Schematic depicting the presence of the three rotational domains of the NNMO film on a cubic STO substrate. }}
\end{figure*}

The octahedral patterns are determined from the presence/absence of certain half-order Bragg peaks, which directly map to the in-phase(+), out-of-phase (-), or absence of tilt (0) of the $B$O$_6$ octahedra, denoted as Glazor’s notations~\cite{Glazer:1972p3384}. Here, we have discussed the 10 uc films on STO and NGO.
NGO and STO substrates exhibit \emph{a$^-$b$^+$c$^-$} and \emph{a$^0$a$^0$a$^0$} patterns, respectively~\cite{Choquette:2016p024105}. For the NNMO films on NGO, the presence of ($\frac{3}{2}$, $\frac{1}{2}$, $\frac{3}{2}$)$_\mathrm{pc}$ Bragg peak satisfying (\emph{H,K,L}) = (odd/2, odd/2, odd/2) ($H$=$L$ $\neq$ $K$), indicates towards the presence of \emph{a$^{-}$/c$^{-}$} rotations (Fig.~\ref{Fig3}(a)). The subsequent absence of (0, $\frac{1}{2}$, $\frac{3}{2}$)$_\mathrm{pc}$ [(\emph{H,K,L}) = (even/2, odd/2, odd/2)] and ($\frac{1}{2}$, $\frac{3}{2}$, 1)$_\mathrm{pc}$ [(\emph{H,K,L}) = (odd/ 2, odd/2, even/2) peaks] as shown in Fig.~\ref{Fig3}(b) affirm to the absence of \emph{a$^{+}$} and \emph{c$^{+}$} rotations. Moreover, ($\frac{1}{2}$, 0, $\frac{3}{2}$)$_\mathrm{pc}$ peak satisfying (\emph{H,K,L}) = (odd/ 2, even/2, odd/2) validates  \emph{b$^+$} tilt [Fig.~\ref{Fig3}(c)].  Hence, the collective results of the half-order diffraction reveal that NNMO film on NGO displays a single domain \emph{a$^-$b$^+$c$^-$} ORP, similar to the ORP pattern of the underlying NGO substrate.

Contrastingly, for the films on STO, we observe additional half-order reflections compared to the films on NGO. Like the previous case, here also we find the ($\frac{3}{2}$, $\frac{1}{2}$, $\frac{3}{2}$)$_\mathrm{pc}$ peak depicting the presence of \emph{a$^-$/c$^-$} rotations [Fig.~\ref{Fig3}(d)]. Furthermore, the ($\frac{1}{2}$, $\frac{3}{2}$, $\frac{3}{2}$)$_\mathrm{pc}$   [(\emph{H,K,L}) = (odd/ 2, odd/2, odd/2) ($K$=$L$ $\neq$ $H$)] and ($\frac{1}{2}$, $\frac{1}{2}$, $\frac{3}{2}$)$_\mathrm{pc}$ [(\emph{H,K,L}) = (odd/ 2, odd/2, odd/2) ($H$=$K$ $\neq$ $L$)] signify the presence of \emph{b$^-$/c$^-$} and \emph{a$^-$/b$^-$} rotations respectively [Fig.~\ref{Fig3}(e),(f))]. The most notable observation lies in the presence of all three possibilities of (\emph{H,K,L}) = (odd/ 2, even/2, odd/2) , (even/ 2, odd/2, odd/2) as well as (odd/ 2, odd/2, even /2) peaks shown in Fig.~\ref{Fig3}(g),(h),(i), namely the (0, $\frac{1}{2}$, $\frac{3}{2}$)$_\mathrm{pc}$, ($\frac{1}{2}$, 0, $\frac{3}{2}$)$_\mathrm{pc}$, ($\frac{3}{2}$, $\frac{1}{2}$, 1)$_\mathrm{pc}$ peaks. These implicate the presence of \emph{a$^{+}$, b$^{+}$} and \emph{c$^{+}$} rotations respectively. Thus, our results unequivocally demonstrate multiple ORP domains : \emph{a$^+$a$^-$c$^-$}, \emph{$a^-$a$^+$c$^-$} and \emph{a$^-$a$^-$c$^+$} [we have denoted both $a$ \& $b$ as $a$ since being equal in-plane lattice parameters rotational angles should have equal magnitude]. As expected for the $a^0a^0a^0$ pattern, all half-order peaks are absent for the STO substrate.  

In Fig.~\ref{Fig3}(j), we have plotted half order peaks with one integer and two unequal half integer indices. These peaks appear only when the integer reciprocal lattice vector is parallel to  the in-phase rotation axis~\cite{Choquette:2016p024105}. 
$A$-site displacements perpendicular to the rotation axis also contribute to the peak intensity~\cite{Choquette:2016p024105, Glazer:1975p756}. Thus, the half order peaks ($\frac{1}{2}$,  $\frac{3}{2}$,  1)$_\mathrm{pc}$, (-$\frac{1}{2}$,  -1,  $\frac{3}{2}$)$_\mathrm{pc}$ and (-1  $\frac{1}{2}$,  $\frac{3}{2}$)$_\mathrm{pc}$  correspond to \emph{a$^-$a$^-$c$^+$}, \emph{a$^-$a$^+$c$^-$} and \emph{a$^+$a$^-$c$^-$}, respectively. The similar peak intensities of the \emph{a$^-$a$^+$c$^-$} and \emph{a$^-$a$^-$c$^+$} ORPs in Fig.~\ref{Fig3}(j) indicate that they are nearly equally populated, while the \emph{a$^+$a$^-$c$^-$} ORP is relatively lower in population. This can be related to the energetics of ORPs on the underlying epitaxial strain~\cite{Sclauzero:2015p235112}. Previous studies have found that perovskite films on cubic substrates typically adopt a mixed $a^-a^+c^-$ and $a^+a^-c^-$ phase under compressive strain, and an $a^-a^-c^+$ phase under high tensile strain ($>$2\%)~\cite{Choquette:2016p024105,Brahlek:2017p4974362,Gunter:2012p214120,Tung:2013p205112}. Surprisingly, the NNMO film exhibits all three ORPs on STO despite moderate tensile strain (schematically depicted in Fig.~\ref{Fig3}(k)).

Perovskite oxides having $Pbnm$ and $P2_1/n$ structures  have antiparallel displacement of $A$-sites in the plane normal to the in-phase rotation axis. Such displacement for the  $a^-a^-c^+$ ORP  give rise to (even/2, even/2, odd/2) peaks~\cite{Ravi:2007p3947,Middey:2018p156801,Bhattacharya:2025p2418490}. The origin of additional half-order peaks highlighted in Fig.~\ref{Fig2}(b) exclusively present in NNMO films on STO is related to this  $a^-a^-c^+$ ORP.

\subsection{Layer-resolved out of plane lattice constant from 1D COBRA}
\begin{figure} [ht!]
	\vspace{-1pt}
	\hspace{0pt}
	\includegraphics[width=0.92\textwidth] {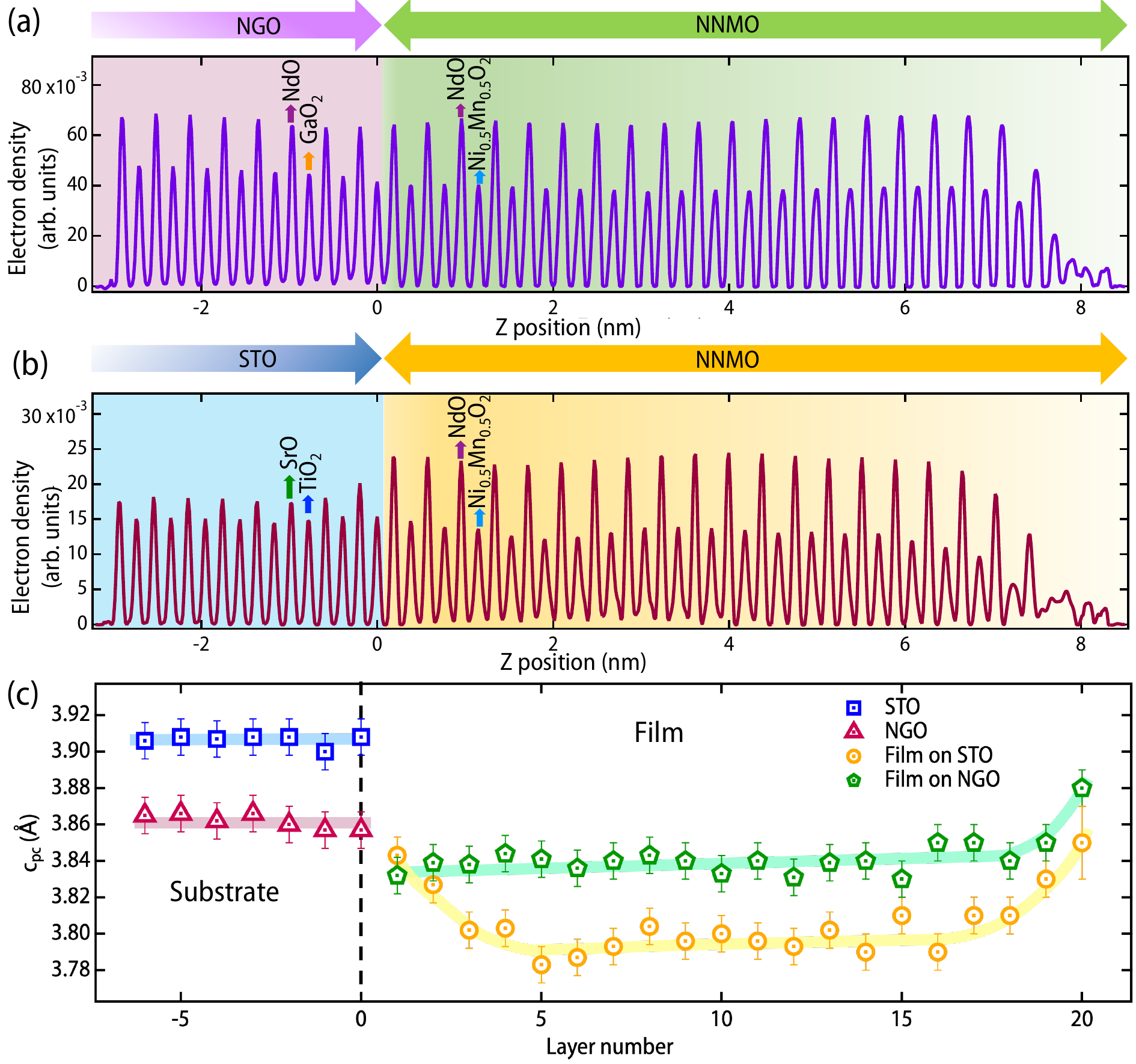}
	\caption{\label{Fig4}{Electron density profile as a function of the out of plane (Z) position for (a) 20 uc NNMO on NGO (b) 20 uc NNMO on STO. (c) Layer-resolved $c_\mathrm{pc}$ of both film and substrate side for 20 uc NNMO on NGO and 20 uc NNMO on STO films. Here '0' of the horizontal axis in all the plots denotes the interface.}}
\end{figure}

In various heterostructures such as LaAlO$_3$/STO, LaMnO$_3$/STO, LaNiO$_3$/STO, La$_{1-x}$Sr$_x$MnO$_3$/ STO with interfacial polarity mismatch,  structural distortions at interfaces and surfaces have been highlighted~\cite{Yu:2014p5118, Kaspar:2019p1801428, Tung:2017p053404, Kumah:2014p1935, Koohfar:2017p024108}.
The diffraction intensity can be expressed as 
\begin{equation}\label{eq1}
 I_c = \frac{1}{2} \eta n \frac{\epsilon_0}{c}| E_c(\vec{q})|^2
\end{equation} 
where $n$ is the refraction index, $\epsilon_0$ is the permittivity of free space and $\eta$ is a detection dependent term related to efficiency, conversion factors, etc. of the detector and $E_c(\vec{q})$ is the scattered wave amplitude from the crystal. Consequently, structural refinement is performed iteratively  by minimization of the crystallographic $R$-factor given by 
\begin{equation}\label{eq2}
 R = \frac{\Sigma |I_{calc} - I_{meas}|}{\Sigma|I_{meas}|} 
 \end{equation} 
where $I_{calc}$ and $I_{meas}$ are the simulated and measured intensities respectively. The inverse Fourier transform of total scattering amplitude yields the electron density map $\rho(\vec{r})$\cite{Disa:2020p1901772, Koohfar:2017p024108}. A user-built MATLAB program has been used to perform the 1D COBRA fitting of (00$L$) Bragg rod and electron density retrieval~\cite{Zhou:2012p195302, Zhou:2010p8103, Yuan:2018p5220} for both 20 uc NNMO on NGO and STO. Best fit (Fig.~\ref{Fig1}(d),(e)) has been obtained following the convergence criteria $R$ $\ll$ 1.

Figure~\ref{Fig4}(a) and (b) show the reconstructed electron density plots with demarcated atomic planes relative to the interface. By measuring the inter-$A$O plane distances, we directly estimate the layer-resolved $c_\mathrm{pc}$ for both substrate and film sides: inter-SrO distances for STO and inter-NdO distances for NGO and NNMO. As shown in Fig.~\ref{Fig4}(c), for the substrate layers, there is no observable change in $c_\mathrm{pc}$ with an average $c_\mathrm{pc}$ value of $\sim$  3.86 {\AA} and 3.91 {\AA} for NGO and STO, respectively. For the film on NGO, the $c_\mathrm{pc}$ remains fairly constant for both the interfacial and bulk-like layers but increases near the surface.  In sharp contrast, the film on STO shows an initial increase of $c_\mathrm{pc}$ for the first 3-4 uc, followed by a decrease to a constant value in the bulk layers. The $c_\mathrm{pc}$ then increases again in the few unit cells near the surface. We further corroborate our findings with a scanning transmission electron microscopy (STEM) study of NNMO/STO, which also revealed a similar trend in $c_\mathrm{pc}$ variation with layer number~\cite{Bhattacharya:2025p176201}. 

Such enhancement in $c_\mathrm{pc}$ has been reported earlier to emanate as a result of some interfacial effects like vanishing tilt due to substrate imprinting ~\cite{Vailionis:2014p105}, and/or cationic intermixing~\cite{Herger:2008p085401}. However, our annular-bright field imaging experiments found that although the octahedral tilt is lower for the interfacial layers compared to the intermediate layers, it is not fully suppressed; rather, the tilt propagates to a few uc on the substrate side itself~\cite{Bhattacharya:2025p176201}. Furthermore, cationic intermixing across the interface is also negligible~\cite{Bhattacharya:2025p176201}. Rather, the correlation with the layer-resolved Mn charge states is employed to interpret the $c_\mathrm{pc}$ trend in the present case. 

Our X-ray absorption spectroscopy (XAS) study confirmed that surface symmetry breaking results in the formation of Mn$^{2+}$ ions near the surface~\cite{Bhattacharya:2025p176201}. The larger ionic radius of Mn$^{2+}$ (0.83 \AA~\cite{Shannon:1976p751}) compared to Mn$^{4+}$ (0.53 \AA~\cite{Shannon:1976p751}) contributes to the enhanced $c_\mathrm{pc}$ at the surface layers for film on both NGO and STO. The interfacial polarity mismatch in NNMO/STO films induces the formation of Mn$^{3+}$ ions (ionic radius $\sim$ 0.645 {\AA}~\cite{Shannon:1976p751}) near the interface, which also results in an increased $c_\mathrm{pc}$ for the initial few unit cells. Overall, we observe a close-knit interplay among electronic reconstruction (charge transfer), structural reconstruction ($c_\mathrm{pc}$ and amount of octahedral distortion) due to polar and structural mismatch across the interface.
 
\subsection{Orbital symmetry from X-ray linear dichroism}

\begin{figure*} [ht!]
	\vspace{-1pt}
	\hspace{0pt}
	\includegraphics[width=0.99\textwidth] {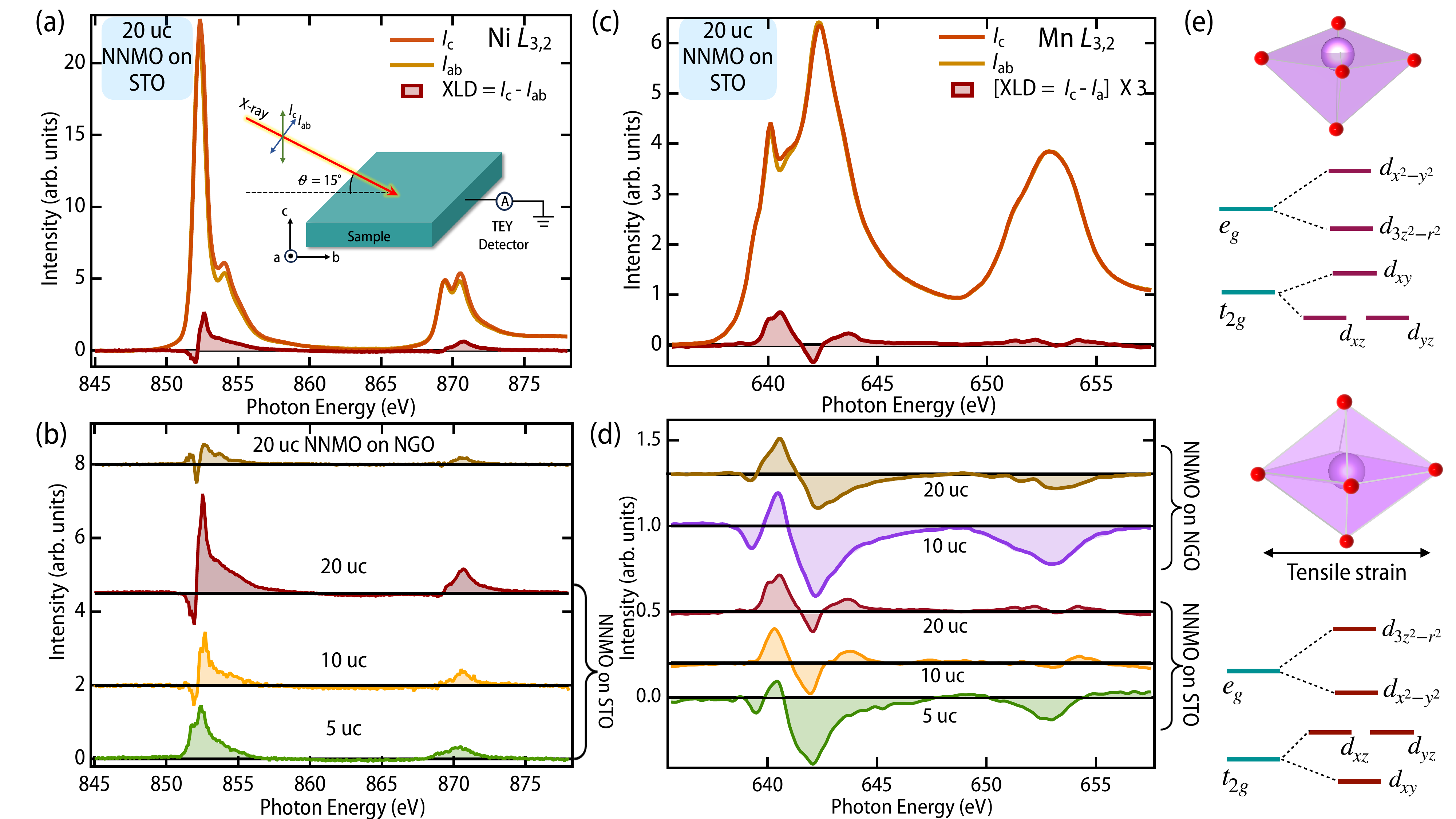}
	\caption{\label{Fig5}{(a) XAS for in-plane polarized X-ray (\emph{I}$_\mathrm{ab}$), out of plane polarized X-ray (\emph{I}$_\mathrm{c}$) and corresponding XLD signal for 20 uc NNMO on STO at (a) Ni $\emph{L}$$_{3,2}$ edge; Inset: Schematic of XAS measurement with linearly polarized light under total electron yield (TEY) mode, \emph{I}$_\mathrm{c}$ denotes vertically polarized spectra and \emph{I}$_\mathrm{ab}$ denotes the horizontally polarized spectra. (c) Mn $\emph{L}$$_{3,2}$ edge. (b) XLD signals at Ni $\emph{L}$$_{3,2}$ edge for  20 uc NNMO on NGO and 5,10 and 20 uc NNMO film on STO. (d) XLD signals at Mn $\emph{L}$$_{3,2}$ edge for  10 and 20 uc NNMO on NGO and 5,10 and 20 uc NNMO film on STO. The XLD spectra have been vertically shifted for visual clarity.(e) MnO$_5$ polyhedron for surface Mn cations under square pyramidal geometry and corresponding crystal field(upper panel). MnO$_6$ octahedron for interfacial Mn cations under tensile strain and corresponding crystal field (lower panel).}}
\end{figure*}

The modification of $c_\mathrm{pc}$ along the growth direction discussed above, rather than the presence of three octahedral rotational patterns, would alter the orbital splitting/symmetry. This must, in turn, affect the magnetic interactions in the NNMO films since the Ni-Mn super-exchange is orbital-dependent~\cite{Das:2008p186402}.
 To investigate the orbital character, we have performed XLD measurements at the Ni and Mn $L_{3,2}$ 
edges (See inset of Fig.~\ref{Fig5}(a) for the experimental geometry)~\cite{Benckiser:2011p189,Green:2021p065004,Pesquera:2012p1189,Chakhalian:2007p318,Middey:2016p056801,Mandal:2023p045145}.
 
The XAS intensity corresponding to the out-of-plane polarized X-ray [denoted by \emph{I}$_\mathrm{c}$ in Fig.~\ref{Fig5}(a),(c)]  and the  in-plane  polarized X-ray [denoted by \emph{I}$_\mathrm{ab}$ in Fig.~\ref{Fig5}(a),(c)]  are directly (inversely) proportional to the hole (electron) occupancies of the $\emph{d}$$_{3z{^2} - r{^2}}$ and  $\emph{d}$$_{x{^2} - y{^2}}$, respectively. Thus, a positive sign of XLD [= \emph{I}$_\mathrm{c}$- \emph{I}$_\mathrm{ab}$]  corresponds to a preferential occupation of the electron in the $\emph{d}$$_{x{^2} - y{^2}}$ orbital and vice versa~\cite{Stohr:2006book}. 
We first discuss results of Ni $L_{3,2}$ edge. XAS measurements [see Fig.~\ref{Fig5}(a)] confirm the presence of Ni$^{2+}$ ($d^8$:$t_{2g}^6$, $e_{g}^2$) in all films, grown on both NGO and STO ~\cite{Bhattacharya:2025p176201}. The sign of XLD is found to be positive, irrespective of the thickness and substrate of the NNMO film,  implying that $\emph{d}$$_{x{^2} - y{^2}}$ orbital is lower in energy than $\emph{d}$$_{3z{^2} - r{^2}}$. This is in concurrence with the tensile strain exerted by the STO substrate [Fig.~\ref{Fig5}(a),(b)]. As shown in Fig.~\ref{Fig5}(b), the XLD magnitude is much smaller for the film on NGO substrate, as expected due to a lower tensile strain (+0.4 \%).

It can be easily anticipated that NNMO films on NGO won't show any orbital polarization for Mn as both Mn$^{2+}$ ($d^5$) and Mn$^{4+}$ ($d^3$) are Jahn-Teller inactive for an octahedral crystal field. However, as shown in Fig.~\ref{Fig5}(d), a prominent negative XLD on Mn $L_2$-edge can be seen for both 10 and 20 uc NNMO films on NGO as supported by finding the integrated area under the XLD signal. This is in sharp contrast even with the expected energy split for tensile strain ($\emph{d}$$_{x{^2} - y{^2}}$ should be lower in energy than $\emph{d}$$_{3z{^2} - r{^2}}$ for in-plane elongation).

Such a surprising lowering of the $d_{3z{^2} - r{^2}}$
  orbital energy can be attributed to Mn$^{m+}$ surface states under square-pyramidal coordination, which arise from the absence of apical oxygen on the film surface. This effect has also been reported in various manganite thin films~\cite{Pesquera:2016p034004, Zenia:2005p024416, Wang:2021p1811}. Such orbital behavior can be further influenced by anisotropic strain response: in-plane and out-of-plane bandwidths may respond differently to strain, potentially leading to inverted orbital polarization with nearly filled bands compared to what is expected from epitaxial strain.

On the contrary, we found a starkly different result in the 10 and 20 uc films on STO. As shown in Fig.~\ref{Fig5}(d), we get a positive XLD for these films, implying preferential occupation in the $\emph{d}$$_{x{^2} - y{^2}}$ orbital.
From the previous discussion for NGO case, we verified that surface states correspond to a negative XLD and bulk Mn$^{4+}$ states would lack any XLD contribution. Hence, we affirm this positive XLD contribution to the Mn$^{3+}$ ($d^4$) species originating due to the polar catastrophe  near the film/substrate interface and that the film is  under tensile strain [lower panel of Fig.~\ref{Fig5}(e)]wherein the interfacial effect on Mn$^{3+}$ cations dominates over the surface symmetry breaking effect on surface Mn cations. Peculiarly, the 5 uc NNMO on STO film depicts a negative XLD similar to the films on NGO. This can be understood by the fact that the surface effect becomes more dominant for a very thin limit~\cite{Calderon:1999p6698,Pesquera:2012p1189}. Overall, our XLD results establish that there exists a thickness-dependent tuning of Mn orbital polarization under the combined effect of polar catastrophe, epitaxial strain and surface effects. 
 
\section{Conclusions} 
In summary, our work demonstrates that coexisting structural and polar symmetry mismatches drive significant lattice and orbital reconstructions in NNMO films on STO. The resulting changes in Mn oxidation state and orbital symmetry directly influence the nature and strength of magnetic interactions~\cite{Das:2008p186402}, providing a compelling explanation for the recently observed thickness-dependent ferromagnetism in NNMO~\cite{Spring:2023p104407}. 
While NNMO films on NGO exhibit a single octahedral rotational domain of $a^-b^+c^-$, films on STO display a complex rotational domain structure, including $a^+a^-c^-$, $a^-a^+c^-$, and $a^-a^-c^+$ domains. Given that the $a^-a^-c^+$ pattern can host ferroelectricity (FE)~\cite{Choquette:2016p024105}, selective stabilization of this domain through epitaxial engineering and geometric lattice manipulation could enable the realization of ferroelectricity in these films~\cite{Liu:2016p133,Kim:2016p68}. This would open up exciting possibilities for designing FM-FE multiferroics~\cite{Hill:2000p6694,Tian:2021p114402}. Furthermore, our study provides a crucial foundation for understanding the orbital physics of our system, particularly in light of recent theoretical predictions of a new orbital Kugel-Khomskii mechanism for stabilizing an FM-FE ground state~\cite{Solovyev:2024p205116}.

\section{Acknowledgement}
The authors acknowledge the use of central facilities of the Department of Physics, IISc, funded through the FIST program of the Department of Science and Technology (DST), Gov. of India. SM acknowledges funding support from a SERB Core Research grant (Grant No. CRG/2022/001906) and from a DST Nano Mission consortium project [DST/NM/TUE/QM-5/2019].  NB acknowledges funding from the Prime Minister’s Research Fellowship (PMRF), MoE, Government of India. JM acknowledges UGC, India for fellowship. This research used resources of the Advanced Photon Source, a U.S. Department of Energy Office of Science User Facility operated by Argonne National Laboratory under Contract No. DE-AC02-06CH11357. This research used resources of the Advanced Light Source, which is a Department of Energy Office of Science User Facility under Contract No. DE-AC02-05CH11231.

\end{document}